\documentclass[12pt,superscriptaddress]{article}
\usepackage{palatino,cite,epsfig,amsmath,amssymb,color}
\def\MeV{\ifmmode {\mathrm{\ Me\kern -0.1em V}}\else
                   \textrm{Me\kern -0.1em V}\fi}
\def\GeV{\ifmmode {\mathrm{\ Ge\kern -0.1em V}}\else
                   \textrm{Ge\kern -0.1em V}\fi}
\def\EE{\mathrm{e^+ e^-}}
\def\EEZH{\mathrm{e^+ e^- \rightarrow ZH}}

\def\zhllqq{\mathrm{ZH\rightarrow \ell^+\ell^- q\bar{q}}}

\def\zhqqqq{\mathrm{ZH\rightarrow q\bar{q} q'\bar{q}'}}
\def\zhllww{\mathrm{ZH}\rightarrow \ell^+\ell^-\mathrm{WW}}
\def\zhqqww{\mathrm{ZH\rightarrow q\bar{q} WW}}
\def\zhllzz{\mathrm{ZH}\rightarrow \ell^+\ell^- \mathrm{ZZ}}
\def\zhqqzz{\mathrm{ZH\rightarrow q\bar{q} ZZ}}

\def\MH{m_{\mathrm{H}}}
\def\MZ{m_{\mathrm{Z}}}
\def\MW{m_{\mathrm{W}}}
\def\ZH{\mathrm{ZH}}
\def\ZZ{\mathrm{ZZ}}
\def\qq{\mathrm{q\bar{q}}}

\def\gg{\mathrm{gg}}

\def\ll{\ell^+\ell^-}
\def\WW{\mathrm{WW}}
\def\Wqq{\mathrm{W \to q\bar{q}^\prime}}
\def\Zqq{\mathrm{Z\to q\bar{q}}}
\def\Hbb{\mathrm{H\to b\bar{b}}}

\def\all{\mathrm{all~final~states}}

\def\EEGGFF{\mathrm{e^+e^-}\rightarrow \gamma^*\gamma^* \mathrm{e^+e^-}\rightarrow \mathrm{f\bar{f}e^+e^-}}
\def\EEWW{\mathrm{e^+e^-\rightarrow W^+W^-}}
\def\EEZZ{\mathrm{e^+e^-\rightarrow ZZ}}
\def\EEZGZG{\mathrm{e^+e^-\rightarrow Z(\gamma^*)Z(\gamma^*)}}

\def\EQQG{\mathrm{e^+e^-\rightarrow q\bar{q} (\gamma)}}

\def\EEWWZ{\mathrm{e^+e^-\rightarrow W^+W^-Z}}
\def\EEZZZ{\mathrm{e^+e^-\rightarrow ZZZ}}
\def\elljet{\rm{2}\ell{\mbox{+2-jet}}}
\def\ellfjet{\rm{2}\ell{\mbox{+4-jet}}}
\def\sjet{\mbox{6-jet}}
\def\fjet{\mbox{4-jet}}
\def\HWW{\mathrm{H\rightarrow WW}}
\def\HZZ{\mathrm{H\rightarrow ZZ}}

\def\ra{\rightarrow}

\begin{document}

%
%
      
\title{
Measurement of the Higgs Boson Mass with a Linear
\mathversion{bold} $\EE$ 
\mathversion{normal}Collider
}

\author{P. Garcia-Abia,\\CIEMAT, Madrid\\[1cm] W. Lohmann and A. Raspereza\\DESY\\}

\maketitle

%
%
\begin{abstract}
The potential of a linear $\EE$ collider operated at 
a centre-of-mass energy of 350$\GeV$ is studied
for the measurement of
the Higgs boson mass. 
An integrated luminosity of 500 fb$^{-1}$ is assumed.
For Higgs boson masses of 120, 150 and 180$\GeV$
the uncertainty
on the Higgs boson mass measurement is estimated to be 40, 65 and
70$\MeV$, respectively. 
The effects of beam related systematics, namely 
a bias in the beam energy measurement, the beam energy spread 
and  the luminosity spectrum due to beamstrahlung, 
on the precision
of the Higgs boson mass measurement are investigated.
In order to keep 
the systematic uncertainty
on the Higgs boson mass well 
below the level of the statistical error, 
the beam energy measurement
must be controlled with a relative precision better
than 10$^{-4}$. 

\end{abstract}


\section{Introduction}

In the Standard Model~\cite{sm}
particles acquire mass due to spontaneous symmetry breaking
by introducing a doublet of complex scalar fields. 
This so called Higgs mechanism~\cite{higgs} leads to one 
scalar particle, the Higgs boson. 
The mass of the Higgs boson is a free parameter of the 
Standard Model and of fundamental nature.
If the Higgs boson exists, the 
Large Hadron Collider at CERN will be able to discover it~\cite{lhc_higgs}. 
Precision measurements of the Higgs boson parameters
and the exploration of the complete Higgs boson profile
will be one of the central tasks at a future linear 
$\EE$ collider.  

In this article we study the potential of a future $\EE$ collider 
for the measurement of the mass of a relatively light Higgs boson, 
in the mass range from 120 to 180$\GeV$,
and investigate 
possible systematic effects influencing the precision of this 
measurement. The analysis presented 
extends previous studies on the measurement 
of the mass of a light Higgs boson~\cite{garcia}
and complements recent studies on the
determination of resonance parameters of the Higgs
boson with the mass in the range from 200 to 320$\GeV$
at a future linear $\EE$ collider~\cite{hhiggs_tesla}.  


\section{Experimental Conditions and Detector Simulations}

The study is performed for a linear collider operated at a centre-of-mass
energy, $\sqrt s$, of 350$\GeV$ and 
an event sample corresponding to an integrated luminosity
of 500 fb$^{-1}$. This integrated luminosity is expected
in about one year of running with design luminosity.

The detector used in the simulation follows the proposal
for the TESLA collider presented in 
the Technical Design Report \cite{tdr}.
The interaction region is surrounded by a central tracker consisting
of a silicon micro-vertex detector as the innermost part
and a time projection chamber. In radial direction follow
an electromagnetic calorimeter, a hadron calorimeter,
the coils of a superconducting magnet
and an instrumented iron flux return yoke.
The solenoidal magnetic field is 4T.
The central tracker momentum resolution is 

\begin{equation}
  \frac{\sigma_{p_t}}{p_t} = 7 \cdot 10^{-5} \cdot p_t~,
\end{equation}
where $p_t$ is the transverse momentum in$\GeV$/c. 

The energy resolutions of the electromagnetic and hadron calorimeters
are:
\begin{equation}
\frac{\sigma_{E_{\rm e}}}{E_{\rm e}} = \frac{10\%}{\sqrt E_{\rm e}} \oplus 0.6 \%,
\hspace{10mm} \frac{\sigma_{E_{\rm h}}}{E_{\rm h}} = \frac{50\%}{\sqrt E_{\rm h}} 
\oplus 4 \%,
\end{equation}
where $E_{\rm e}$ and $E_{\rm h}$ are the energies
of electrons and hadrons in$\GeV$.
The polar
 angular coverage  of the central tracker maintaining the resolution is
$|\cos \theta| < 0.85$, above this range the tracking resolution deteriorates.
The electromagnetic and hadron calorimeters 
cover $|\cos \theta| < 0.996$ 
maintaining the resolution over the whole angular range.
The simulation of the detector is done using
the SIMDET \cite{simdet} package. 

The event reconstruction is done in terms of particle flow objects. 
First, tracks are measured with the tracking system and associated 
to calorimeter clusters to define charged particle flow objects of
electrons, muons and charged hadrons. Since the momentum 
measurement
by the tracking system is much more accurate than the angular and energy
measurements with calorimeters, the tracking information
is used for 
the determination of the four-momentum of charged particles.  
Calorimetric clusters with no associated track are regarded as 
neutral particle flow objects originating from photons and neutral hadrons. 
Measurements of the four-momentum of neutral objects are solely
based on the calorimetric information.

\section{Physics Processes}

At a centre-of-mass energy of $\sqrt s $ = 350$\GeV$, 
the dominant process for 
light Higgs boson production in the Standard Model is $\EEZH$.
Events of this process, hereafter referred to as signal, 
are generated using PYTHIA
\cite{pythia}
for Higgs boson masses, $\MH$, of 120$\GeV$, 150$\GeV$ and 180$\GeV$.
For the Higgs boson, all decay modes are simulated
as expected in the Standard Model. The decay modes into hadrons, $\WW$
and $\ZZ$ are investigated in detail.
Z decays are considered into electrons, muons and hadrons.
The Standard Model cross sections 
are given in Table~\ref{hevents} 
for the investigated signal channels.
\begin{table*}[ht]
\begin{center}
\begin{tabular}{|l|c|c|c|c|}
\hline 
   decay mode             & topology         & \multicolumn{3}{c|}{cross section, fb}                        \\ 
                          &          &120$\GeV$  &  150$\GeV$ &      180$\GeV$ \\ 
\hline
   $\ZH\to\ll\qq, \ZH\to\ll\gg$ & 2$\ell$+2-jets  &   8.8          &   2.0               &   0.06               \\
   $\ZH\to\qq {\rm q'\bar{q}'}, \ZH\to\qq\gg$& 4-jets &  91.9      &  20.4               &   0.62               \\
   $\ZH\to\ll\WW, \ \Wqq$ & 2$\ell$+4-jets   &   0.6               &   2.6               &   2.6               \\
   $\ZH\to\qq\WW, \ \Wqq$ & 6-jets           &   6.0               &  26.5               &  26.6               \\ 
   $\ZH\to\ll\ZZ, \ \Zqq$ & 2$\ell$+4-jets   &  0.08               &  0.33               &  0.17               \\
   $\ZH\to\qq\ZZ, \ \Zqq$ & 6-jets           &  0.82               &  3.46               &  1.73               \\ 
\hline
   $\ZH\to\all$           &                  & 160.3               & 123.7               &  89.0               \\
\hline
\end{tabular}
\caption{The  cross  sections, in fb, times the  branching fractions 
         of the investigated
         signal  final  states  for Higgs  boson  masses of 120$\GeV$, 150$\GeV$ and
         180$\GeV$ 
         as predicted in the Standard Model for
         $\sqrt s =$ 350$\GeV$. 
         Also given is the total cross section for $\EEZH$. 
         The cross sections
         are calculated with PYTHIA taking into account
         initial state radiation.
\label{hevents}}
\end{center}
\end{table*}

For background estimations 
events are generated with PYTHIA for
the processes\hspace{1mm} $\EQQG$, $\EEWW$, $\EEZGZG$
and $\EEGGFF$.
Six fermion final states resulting 
from the triple gauge boson production are generated with 
the WHIZARD package~\cite{whizard}. 
The cross sections of the main background reactions 
are given in Table~\ref{table:bevents}.
\begin{table}[ht]
\begin{center}
\begin{tabular}{|l|c|c|}
 \hline
background process    & cross section, fb & events   \\ 
 \hline
 $\EEGGFF$  & $4.0\times 10^6$ & $2.0\times 10^9$    \\
 $\EQQG$    & $2.7\times 10^4$ & $1.4\times 10^7$    \\
 $\EEWW$    & $1.3\times 10^4$ & $6.5\times 10^6$    \\
 $\EEZGZG$  & $1.0\times 10^3$ & $5.0\times 10^5$    \\ 
 $\EEWWZ$   &      $13.2$      & $6.5\times 10^3$    \\
 $\EEZZZ$   &      $0.48$      & $9.6\times 10^2$    \\
\hline
\end{tabular}
\caption{
         The cross sections, in fb, and
         the  numbers of events expected for the important
         background processes at a centre-of-mass energy 350$\GeV$. 
         An integrated luminosity of 
         500 fb$^{-1}$ 
         is assumed.
\label{table:bevents}
}
\end{center}
\end{table}
The numbers of events generated for each background channel
as well as the number of generated signal events correspond
to an integrated luminosity of 500 fb$^{-1}$.

Initial state radiation is simulated by PYTHIA. Beamstrahlung
is taken into account using the CIRCE program \cite{circe}.

\section{Analysis Procedure}

The measurement of the Higgs boson mass 
is based on the reconstruction of the exclusive final 
states $\zhllqq$, $\zhqqqq$, $\zhllww$ and $\zhqqww$. 
In the latter two cases, the contributions from the $\zhllzz$ and $\zhqqzz$
final states are also taken into account.

The analysis in all channels proceeds as follows. 
First, a selection of events of the specific topology is applied 
to the samples of signal and background events
exploiting event shape variables
and lepton identification.  
For Higgs boson masses below 150$\GeV$, the decay 
$\Hbb$ is dominant, leading to $\elljet$
and $\fjet$ topologies. For larger 
Higgs boson masses, the $\HWW$
decay becomes dominant, leading to  
$\ellfjet$ and $\sjet$ final states. 
For their study, jet identification is crucial.
A kinematic fit imposing energy and momentum conservation
improves considerably the di-jet or 4-jet mass resolutions,
and hence the accuracy of the mass measurement.
In order to construct the covariance matrix used in the kinematic fit,
the resolutions of the lepton and jet energies and angular
measurements are needed. For the lepton momentum, the tracker resolution 
from Equation (1) is used. The resolution in the polar and 
azimuthal angles, $\theta$ and $\phi$, of the lepton momentum vector are obtained 
from Monte Carlo as: 
\begin{equation}
\sigma_\theta = 1~\mathrm{mrad},\hspace{10mm}
\sigma_\phi = \frac{\sigma_\theta}{\sin\theta}.
\end{equation}
The resolutions of the jet energies and angular measurements
are obtained from a Monte Carlo study using the sub-detector
resolutions
from Equations (1) and (2). They are parameterized as:
\begin{equation}
\sigma_E/E = \frac{30\%}{\sqrt{E}}, 
\hspace{5mm} \sigma_\theta = 15~\mathrm{mrad}, 
\hspace{5mm} \sigma_\phi = \frac{\sigma_\theta}{\sin\theta},
\end{equation} 
where $E$ is the energy, $\theta$ the polar and $\phi$ the azimuthal
angle of the jet.

The Higgs boson mass is obtained  
by fitting the invariant mass 
spectrum of the jets assigned to the Higgs boson decay
with a superposition of signal 
and background distributions. The shape of the signal distribution
is derived from a high statistics Monte Carlo sample of signal events
and kept fix in the fit.
Free parameters are the peak value
and the normalisation factor of the signal distribution.

\subsection{The $\mathbf{\zhllqq}$ and $\mathbf{\zhqqqq}$~ Final States}

These final states are characterized by two isolated leptons
and two jets or by four jets and have the full energy deposited in the detector. 
Hence, events where  
the total energy visible in the detector is less 
that 80\% of the centre-of-mass energy are rejected.
Global event characteristics are used for the signal selection. 

For the channel $\zhllqq$, the
number of reconstructed particles must be greater than 20, the event thrust, $T$, 
must be less than 0.85 and the
absolute value of the cosine of the
 polar angle of the thrust vector, $\cos \theta_T$, 
must be less than 0.9.
Electrons are identified as energy deposits in the electromagnetic
calorimeter whose shape is compatible with the expectation for an 
electromagnetic shower and with a matched track in the central tracker.
The measured track momentum and shower energy must be in agreement
within 5\% 
and the shower leakage into the hadron calorimeter
must be less than 2$\GeV$. 
Muons are tracks pointing to energy deposits in the calorimeters
which are consistent with
the expectation for a minimum ionizing particle. 
A pair of electrons or muons
with opposite charge is required. 
Both electrons and muons must have momenta larger than 10$\GeV$/c and 
fulfill the polar angle cut $|\cos \theta_\ell|$ $<$ 0.9.
Leptons must satisfy isolation criteria, meaning that there are 
no other particles reconstructed within a 15$^o$ cone with respect  
to the lepton momentum vector. The invariant mass of a pair of leptons
must be compatible with the mass of the $\mathrm{Z}$ boson within 10$\GeV$.
These criteria reduce the backgrounds listed in
Table~\ref{table:bevents} in the selected sample to the level 
of a few \% with the exception of the process
$\EEZZ$. 
A cut on the polar angle of the momentum vector of
the di-electron or di-muon system,
$|\cos\theta_{\ell\ell}|$ $<$ 0.9, further suppresses the $\mathrm{ZZ}$ background.
The signal selection efficiency is about 45\%.
All reconstructed particles, except the two isolated leptons,
are
grouped into two jets using the Durham~\cite{durham}
jet clustering algorithm. 

Event selection for the  $\zhqqqq$ channel
is performed by requiring
the number of reconstructed particles
to
be larger than 40, $T <$ 0.85  and 
$|\cos \theta_T| <$ 0.8.
No isolated leptons with an 
energy greater than 10$\GeV$ are allowed.
Reconstructed particles
are grouped
into four jets using the
Durham
jet clustering algorithm. 
Events are retained if
the jet resolution parameter, for which the event 
is resolved from the four- to three-jet topology, $y_{34}$,
fulfill the relation log $(y_{34}) > - $5.

The selected events of both final states  
are subject to a kinematic fit~\cite{blobel} imposing energy
and momentum conservation. The kinematic fit is performed by
varying the lepton momenta and angles 
within their resolutions given by Equations (1) and (3), respectively.
The jet energies and angles are varied within the corresponding
resolutions given by Equations (4). 

\begin{figure}
\includegraphics[width=0.7\textwidth]{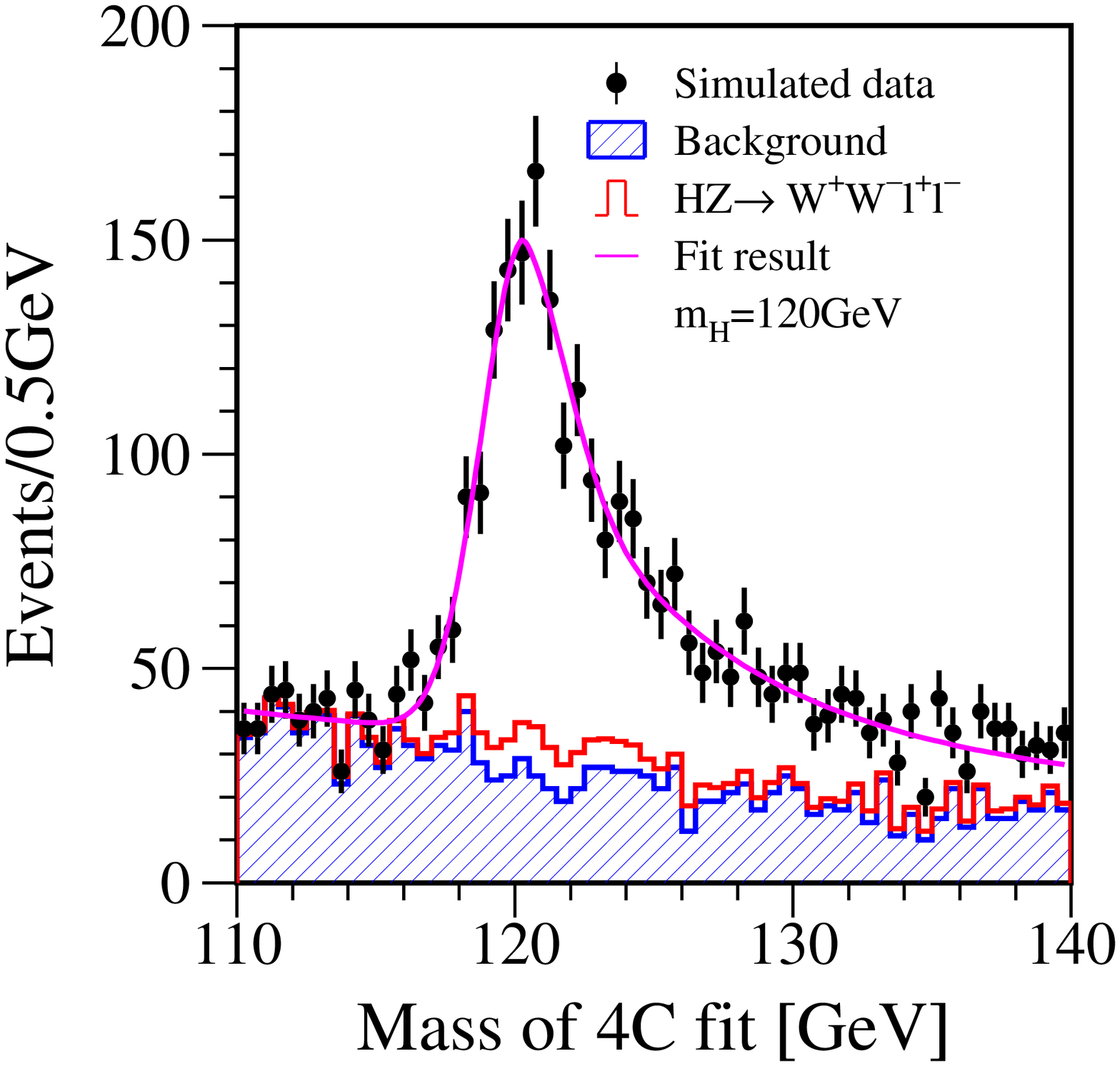}
\includegraphics[width=0.7\textwidth]{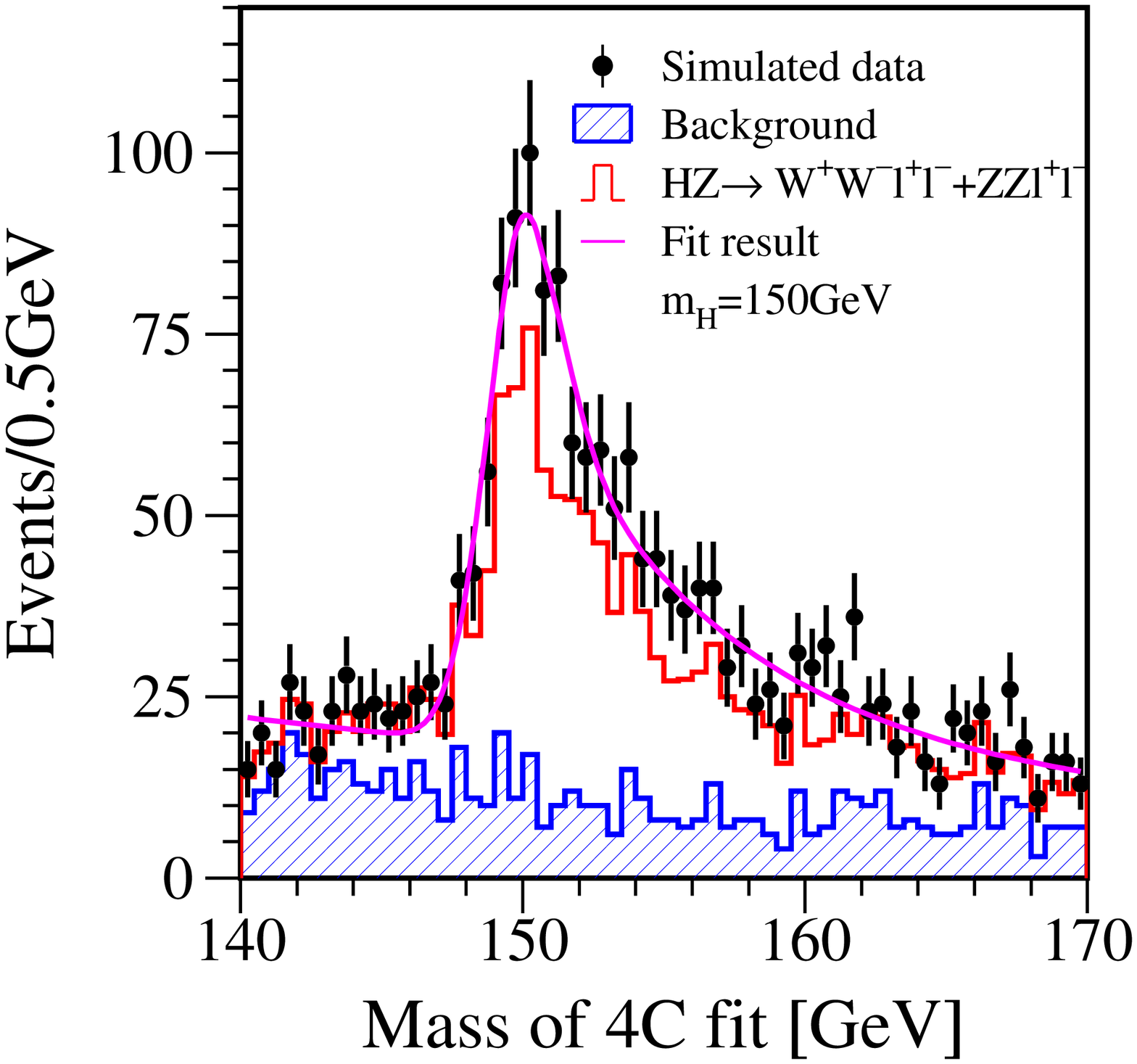}
\caption[]{\label{fig:4Cmass120} The di-jet invariant mass 
from the $\zhllqq$ final
state after a 4C kinematic fit for $\MH$ = 120$\GeV$ (top)
and 150$\GeV$ (bottom).}
\end{figure}

For
events selected as $\zhllqq$, energy and momentum conservation
results in four constraints (4C fit).
Since the experimental resolution in the invariant mass of the di-lepton 
system is much smaller than the natural width of the $\mathrm{Z}$ boson, no constraint 
is applied in the kinematic fit to force the di-lepton mass to $\MZ$. 
The di-jet invariant mass spectra after the 4C fit
are shown in 
Figure~\ref{fig:4Cmass120}
for $\MH$ = 120$\GeV$ and 150$\GeV$, respectively.
Clear signals are seen on top of the remaining smooth background from
$\EEZZ$. 
Also shown are the contributions from $\HWW$ and $\HZZ$ decays to the signal.
These are negligible for $\MH$ = 120$\GeV$ but amount to 62\% and 5\%, respectively, of the signal
for $\MH$ = 150$\GeV$.  

The masses obtained from the fits equal the generated Higgs boson masses and 
have errors of 85$\MeV$ for $\MH$ = 120$\GeV$
and 100$\MeV$ for  $\MH$ = 150$\GeV$. 

\begin{figure}
  \includegraphics[width=0.7\textwidth]{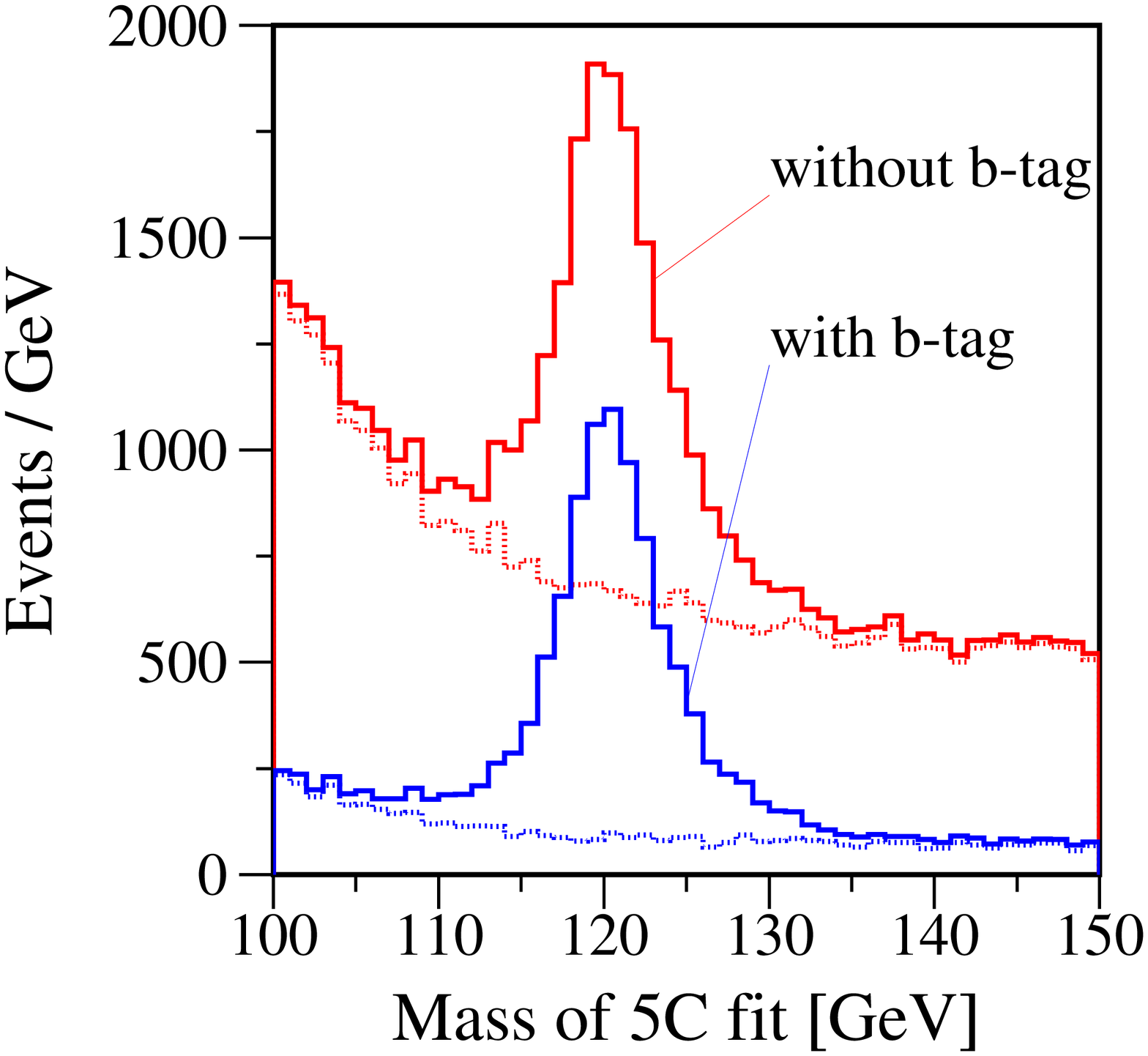}
  \caption[]{\label{fig:b-tag_HZqqqq}
The distribution of the invariant mass 
of the two jets assigned to the Higgs boson decay in the
$\zhqqqq$ final state
without requirement on the b-tag and after requiring the
values of the b-tag of two jets to be larger than 0.2. 
    }
\includegraphics[width=0.7\textwidth]{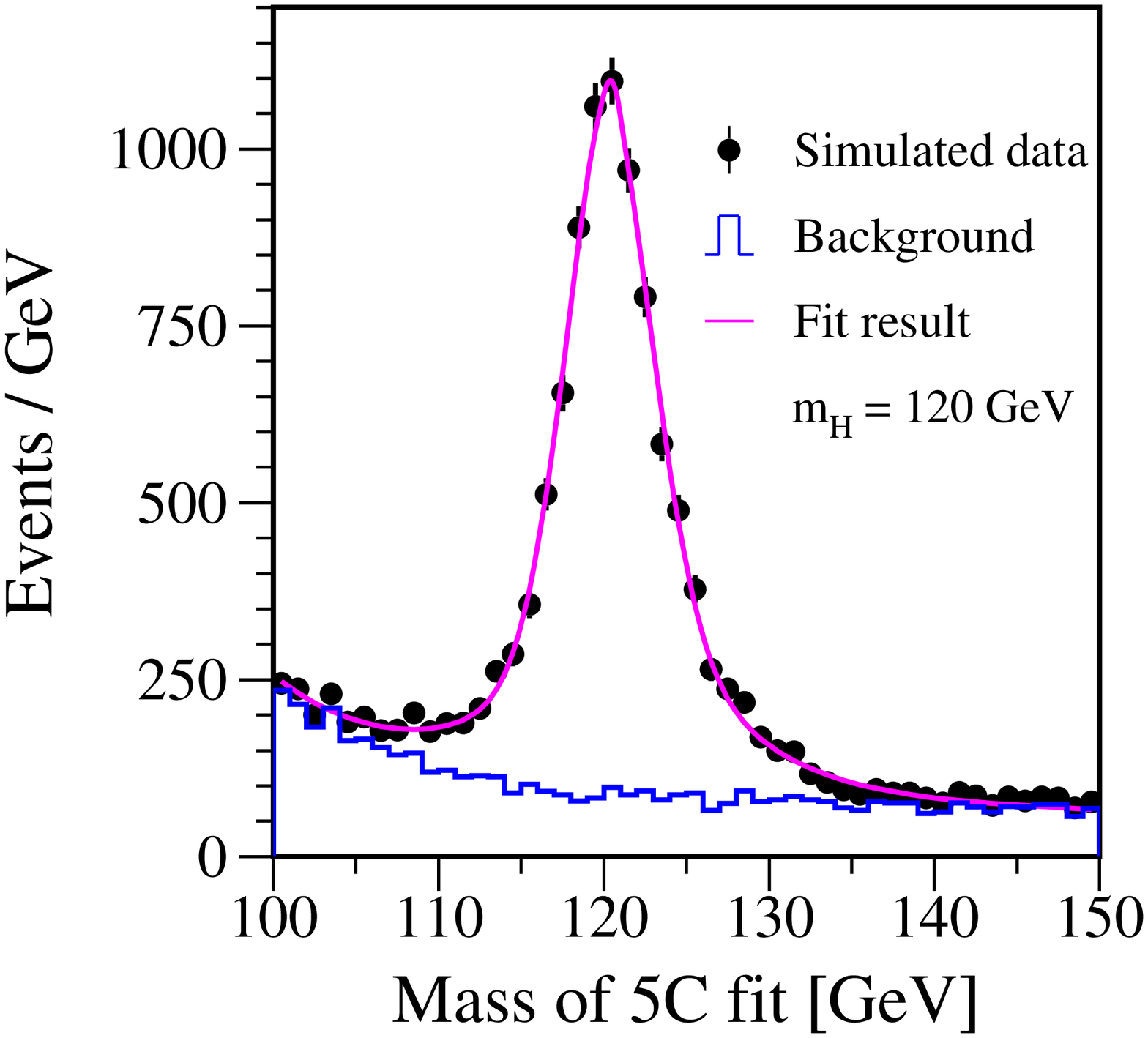}
\caption[]{\label{fig:5Cmass120}
The invariant mass
of the two jets assigned to the Higgs boson decay
in the $\zhqqqq$ final
state after the 5C kinematic fit for $\MH$ = 120$\GeV$.
For two jets the  b-tag must be larger than 0.2.
}
\end{figure}

For the $\fjet$ final states, in addition to the four constraints
from energy and momentum conservation,
the invariant mass of the two jets 
assigned to the $\mathrm{Z}$ boson decay is constrained to $\MZ$.
Hence, a 5C fit is performed for all possible di-jet pairings.
The pairing with the minimal 
 $\chi^2$ is chosen. 
In addition, this $\chi^2$ must be less than 70. 
The signal selection efficiency is about 25\%, however
the remaining event sample contains considerable background
from $\EEZZ$, $\EEWW$ and $\EQQG$. 
The signal-to-background ratio 
is enhanced using the identification of
b-quark jets. The ZVTOP~\cite{kuhl} topological
vertex finder adapted for the pixel micro-vertex
detector~\cite{tdr} is used to search for secondary 
vertices inside jets and determine mass, momentum and decay length
of the vertex.
In addition, the impact parameter joint probability~\cite{barate}
and the two highest impact parameter significances are used   
as input into neural networks trained with jets containing no, one and 
more than one
secondary vertices.
A jet b-tag variable is defined~\cite{desch} as function  
of the neural network output $x$ as
\begin{displaymath}
B(x) =  \frac{f_b(x)}{f_b(x)+f_{udsc}(x)},
\end{displaymath}
where $f_b$ and $f_{udsc}$ are probability density functions
of the neural network outputs in samples of b-jets and udsc-jets, respectively. 
The improvement in the signal to background ratio can be seen
in Figure~\ref{fig:b-tag_HZqqqq}, where invariant mass distributions
are shown for the two jets assigned to the Higgs boson decay
without a requirement on the  
jet b-tag variable and for events with 
at least two jets satisfying $B(x) >$ 0.2. The latter requirement keeps the 
signal statistics almost unchanged but reduces significantly the background.
For $\MH$ = 150$\GeV$ also events from  $\zhqqzz$ and $\zhqqww$ are selected
amounting to 25\% of the signal.
Using the invariant mass distributions 
obtained with the requirement
$B(x) >$ 0.2
the fit of the 
Higgs boson mass is performed.
As an example,
the di-jet invariant mass distribution and the fitted function of the signal 
is shown in Figure~\ref{fig:5Cmass120}. 
   
The results for the Higgs boson masses are equal to the generated masses.
The statistical 
errors are 45$\MeV$ at $\MH$ =120$\GeV$
and 170$\MeV$ at  $\MH$ = 150$\GeV$.

\subsection{The $\zhllww$ and $\zhqqww$~ Final States}

We consider $\mathrm{W}$-boson decays into two quarks, hence the
topologies of these final states are two isolated leptons accompanied
by four jets or six jets, respectively. 
The requirements for electron and muon identification
are the same as in the previous section. Although event selection  
is optimized specifically for the $\zhllww$ and $\zhqqww$ final states,
contributions from the $\zhllzz$ and $\zhqqzz$ channels are also 
taken into account.

Events are selected 
with an
energy
deposited in the detector of more 
  than 80\%
of the centre-of-mass energy and
a number 
of the reconstructed particles
larger than 40. 
 
Events of the final state
$\zhllww$ must contain a pair
of isolated electrons or muons with opposite charges.
Furthermore, the event 
thrust and the polar angle of the thrust
vector   
are used to suppress the dominant background from
the $\mathrm{WW}$ and $\mathrm{ZZ}$ final states. The values of the cuts 
are $T$ $<$ 0.95 and $|\cos\theta_T|$ $<$ 0.95.  
Since the two leptons
of $\zhllww$ originate from the $\mathrm{Z}$ decay, their invariant mass
is required
to be equal within 10$\GeV$ to $\MZ$. 
A cut on the polar angle
of the di-lepton momentum vector, $|\cos\theta_{\ell\ell}|$ $<$ 0.9, 
further suppresses the $\mathrm{ZZ}$ background.
Tracks  and calorimetric energy deposits
not stemming from the leptons are grouped
into four jets
using the Durham algorithm. 
The jet resolution parameter $y_{34}$
must satisfy log$(y_{34})$ $>$ $-$6.0.

\begin{figure}
\includegraphics[width=0.7\textwidth]{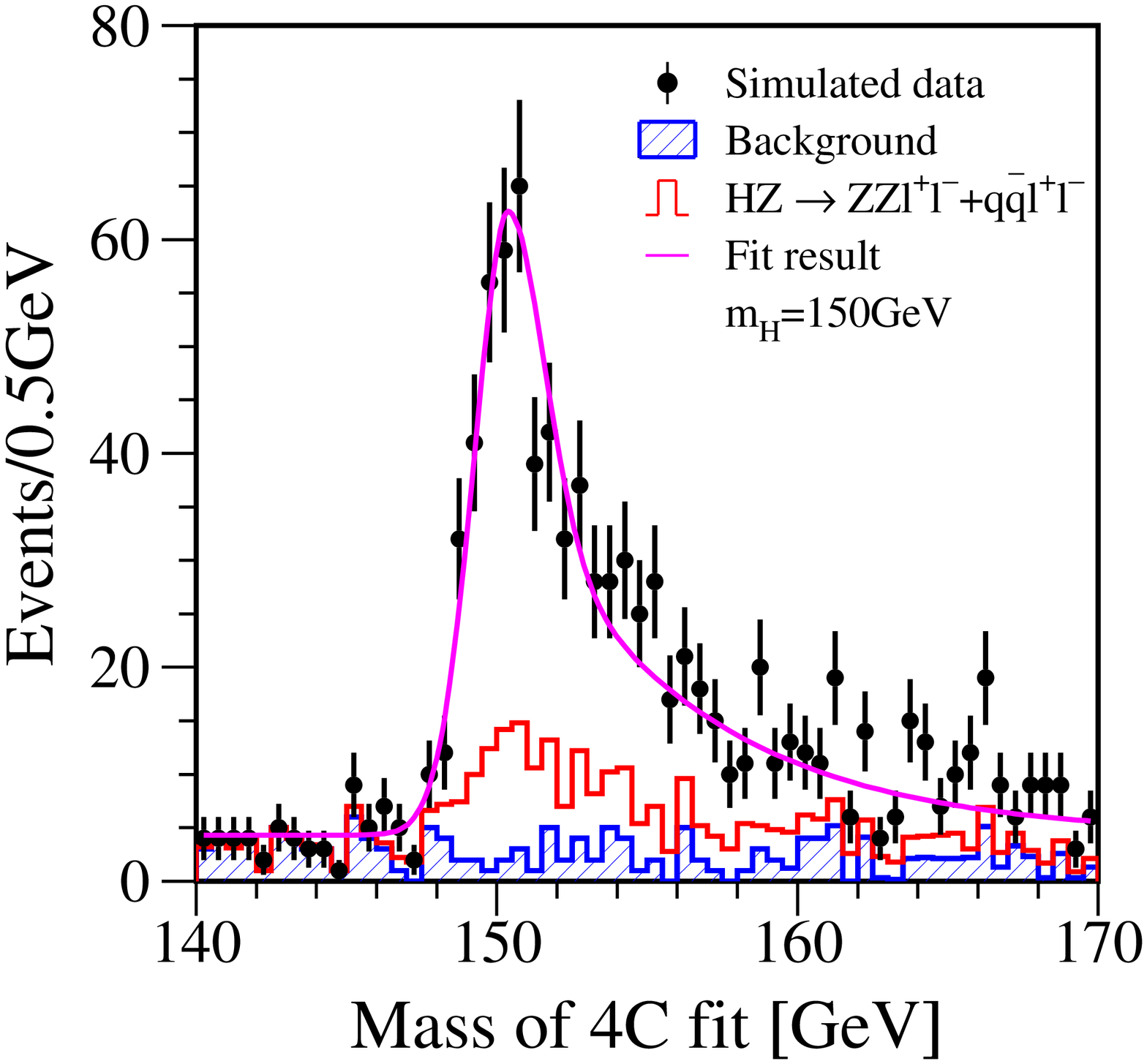}
\includegraphics[width=0.7\textwidth]{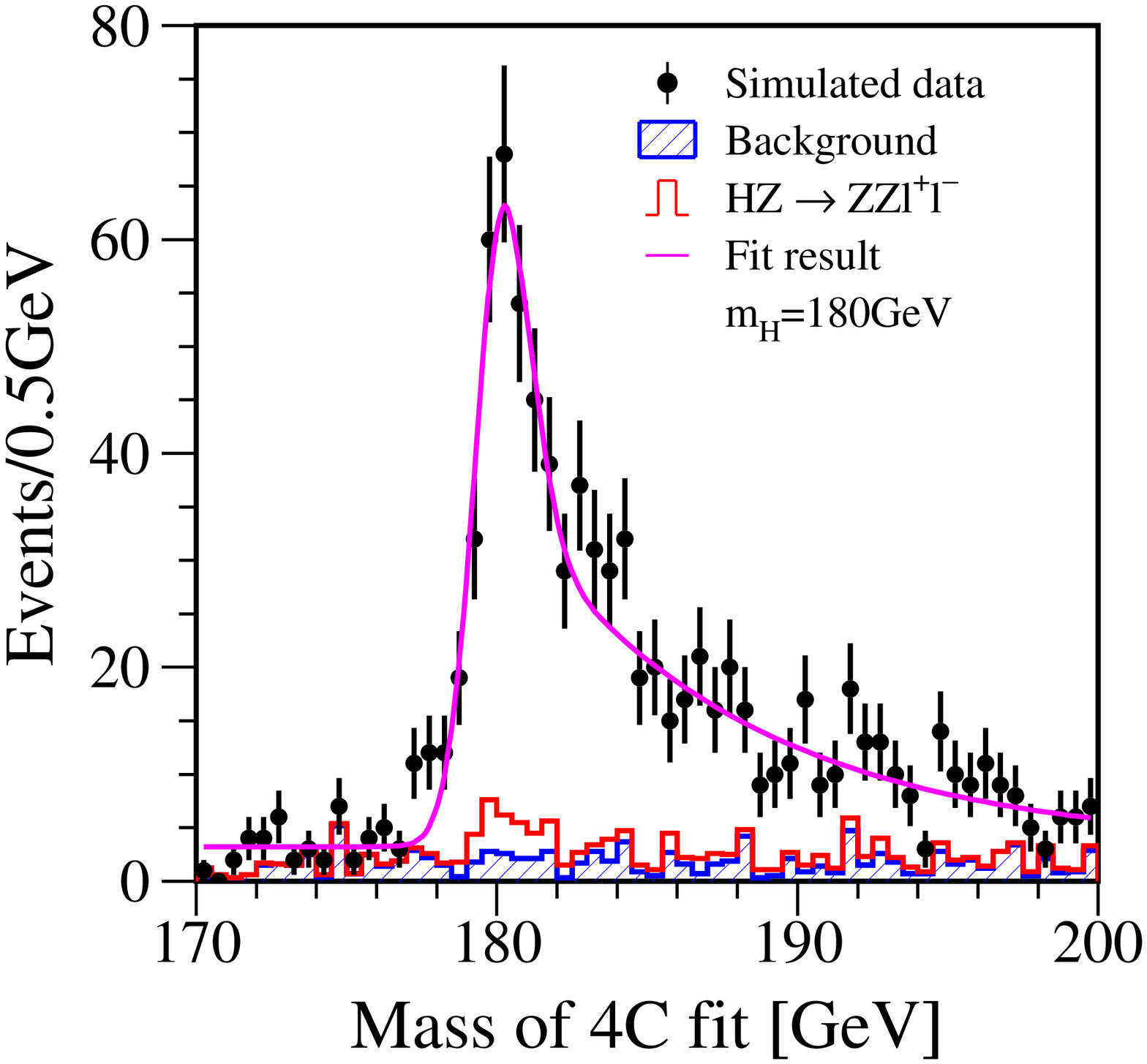}
\caption[]{\label{fig:4C4jetmass} The $\fjet$ invariant mass 
from the $\zhllww$ final
state after a 4C kinematic fit for $\MH$ = 150$\GeV$ (top)
and 180$\GeV$ (bottom).}
\end{figure}

Then a 4C kinematic fit is performed
imposing energy and momentum conservation.
Only events for which the $\chi^2$ of the 4C fit is less than 50
are retained in the selected sample. The signal selection efficiency amounts to 50\% 
at $\MH =$ 150$\GeV$ and 60\% at $\MH =$ 180$\GeV$.
The $\fjet$ invariant mass distributions
after the kinematic fit are shown in 
Figure~\ref{fig:4C4jetmass}
for $\MH$ = 150$\GeV$ and 180$\GeV$.

The fit of the mass spectra 
in Figure~\ref{fig:4C4jetmass} again results in mass values for the Higgs boson
equal to the generated ones.
The uncertainties of the masses amount to 90 and 80$\MeV$
for $\MH$ = 150$\GeV$ and $\MH$ = 180$\GeV$, respectively. 

The small background in this channel comes mainly from the semileptonic 
decays of pair produced $\mathrm{Z}$ bosons and triple gauge boson  
production, $\mathrm{ZWW}$, with a leptonic $\mathrm{Z}$ decay.
Events of the process $\rm ZH\to \ell^+\ell^-$ $\mathrm{ZZ}$ 
constitute 13\%  and 6\% of the signal
in the selected sample for $\MH$ = 150 and 180$\GeV$, respectively.
    
\begin{figure}
\includegraphics[width=0.7\textwidth]{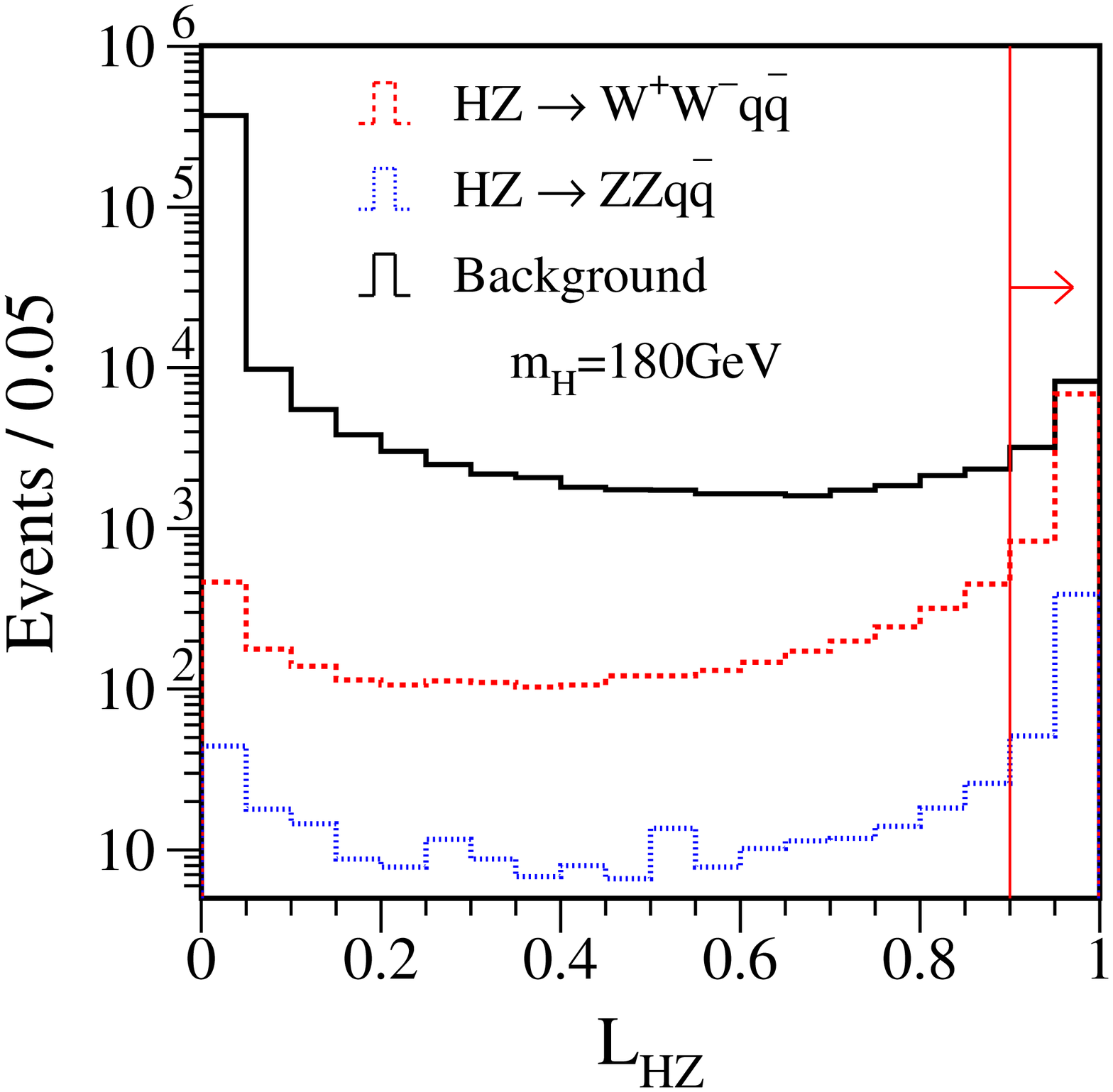}
\caption[]{\label{fig:cuts6q} 
The distributions of the signal likelihood used
to select $\HWW \ra \sjet$ final states for
$\MH$ = 180$\GeV$.
Solid, dashed and dotted 
lines represent 
the background processes, 
$\zhqqww$ and $\zhqqzz$
signals. The
vertical line indicates the cut imposed on this quantity.
}
\end{figure}

The $\zhqqww$ channel is selected by 
requiring 
$T < $ 0.9 and
$|\cos\theta_T|$ $<$ 0.95. 
There must be no isolated leptons 
with an energy greater than 10$\GeV$. 
The reconstructed particles are grouped into six jets using
Durham jet algorithm. The
jet resolution parameter, 
for which an event is resolved from the 6- to 5-jet topology, $y_{56}$, 
must satisfy 
log$(y_{56}) > -$8.
Then a likelihood discriminant, L$_{\rm HZ}$, is defined 
using as input the number of particles reconstructed in an event, 
the polar angle of the thrust vector and the
jet resolution parameters $y_{34}$ and $y_{56}$.
Events are accepted when the
value of this discriminant is larger than 0.9.
As an example, Figure~\ref{fig:cuts6q} shows the distribution of 
L$_{\rm HZ}$
for the signal events for $\MH$ = 180$\GeV$ 
and the background processes.  
The six jets are now grouped in three di-jet pairs following criteria
which depend on the mass of the Higgs boson.
For $\MH < 2\MW$ usually only one $\mathrm{W}$ is expected to be on the mass shell, while
the other is produced with a mass close to the difference
between $\MH$ and $\MW$.
The quantity
\begin{displaymath}
\chi^2 = (m_{\rm {ij}}-\MZ)^2/\sigma_{\rm Z}^2 + ( m_{\rm{kl}}-\MW)^2/\sigma_{\rm W}^2
+ (m_{\rm {mn}}-m_{\rm {klmn}}+\MW)^2/\sigma_{\rm W^*}^2
\end{displaymath}
is calculated for all possible di-jet combinations,
where $m_{\rm{ij}}$ is the invariant mass of
the two jets
assigned to the $\mathrm{Z}$ boson, $m_{\rm{kl}}$ the invariant mass of two jets
assigned to the on-shell $\mathrm{W}$ boson, $m_{\rm{mn}}$ the invariant mass of two jets 
assigned to the off-shell $\mathrm{W}$ boson and $m_{\rm{klmn}}$ the invariant mass
of the four jets assigned to decay $\rm H\ra \rm WW^*$. 
The quantities $\sigma_{\rm Z}^2$, $\sigma_{\rm W}^2$ and $\sigma_{\rm W^*}^2$
are obtained from Monte Carlo studies as the convolution
of the bosonic widths and the mass resolutions and
are estimated to be 6, 9 and 15$\GeV$, respectively.  
For $\MH > 2\MW$ both $\mathrm{W}$ bosons are on shell.
Hence all di-jet combinations are taken and the quantity
\begin{displaymath}
\chi^2 = (m_{\rm {ij}}-\MZ)^2/\sigma_{\rm Z}^2 + (m_{\rm {kl}}-\MW)^2/\sigma_{\rm W}^2
+ (m_{\rm{mn}}-\MW)^2/\sigma_{\rm W}^2
\end{displaymath}  
is calculated.
The jet pairing with the smallest value of $\chi^2$ is chosen and subject of
a kinematic fit imposing energy-momentum conservation
and constraining the mass of the two jets assigned to the $\mathrm{Z}$ boson
to $\MZ$. Events are selected into the final sample if the
$\chi^2$ of the 5C fit is less than 30. 
In addition, the fitted mass of the jets originating from 
the on-shell $\mathrm{W}$ decay must be equal to $\MW$ within 20$\GeV$ in the event sample 
selected for 
$\MH$ = 150$\GeV$. For $\MH$ = 180$\GeV$,  the sum and the difference of the fitted masses
of the two jet pairs assigned to a    
W decay must be between 125$\GeV$ and 185$\GeV$ and $-$20 and 20$\GeV$, respectively.
The signal selection efficiency amounts to about 20\%. The sample selected for
$\MH$ = 150$\GeV$ also contains 5\% signal from the $\mathbf{\zhqqqq}$ final state.

\begin{figure}
\includegraphics[width=0.7\textwidth]{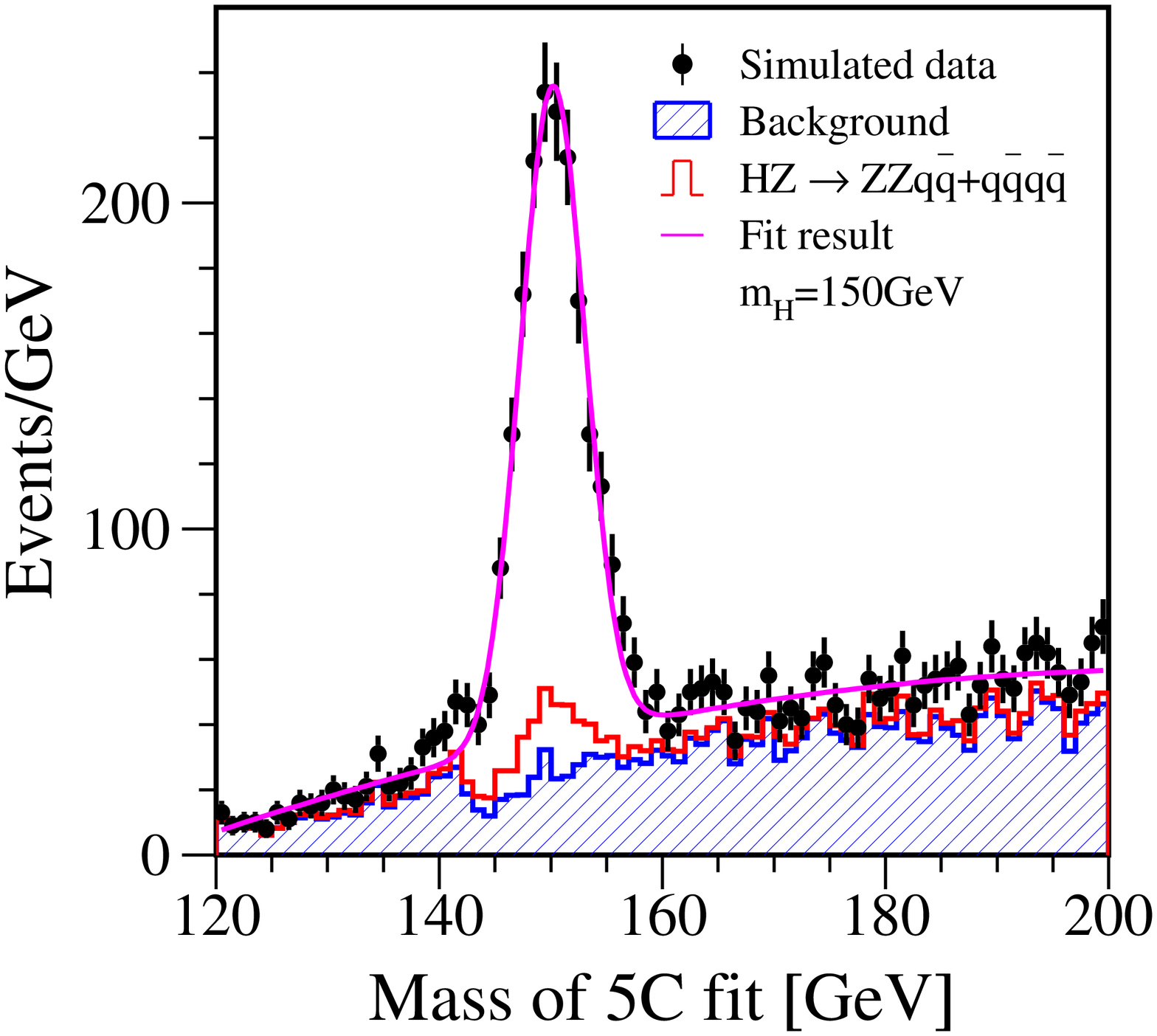}
\includegraphics[width=0.7\textwidth]{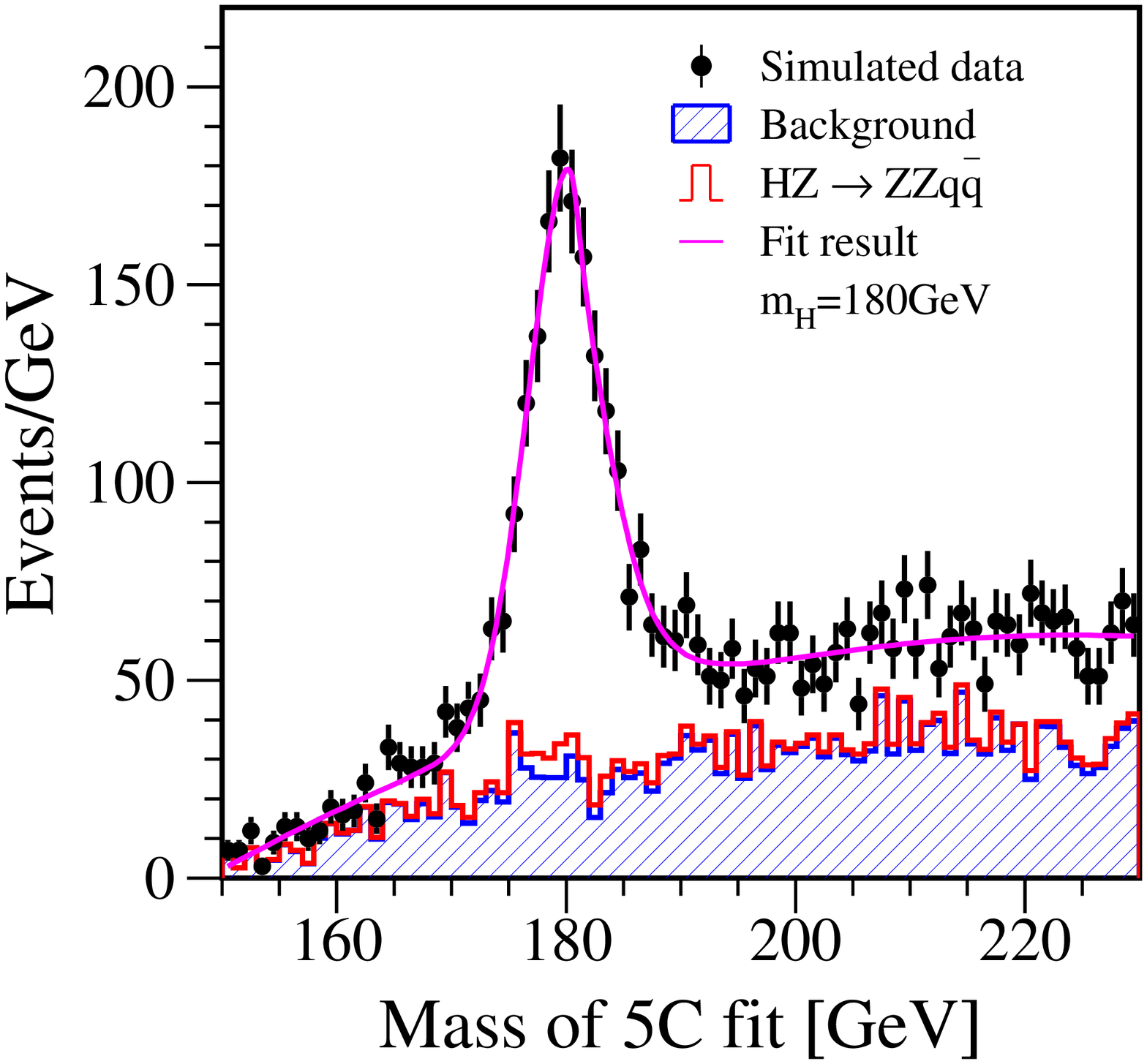}
\caption[]{\label{fig:5C4jetmass} The four jet invariant mass 
from the $\zhqqww$ final
state after a 5C kinematic fit for $\MH$ = 150$\GeV$ (top)
and $\MH$ = 180$\GeV$ (bottom).}
\end{figure}

The distribution of the invariant mass of the $\fjet$ system
is shown in Figure~\ref{fig:5C4jetmass}
for $\MH$ = 150$\GeV$ and 180$\GeV$, respectively.
From the fit approximating the signal by a Gaussian
the uncertainties of the masses are 100$\MeV$ and 150$\MeV$ for 
$\MH$ = 150$\GeV$ and 180$\GeV$, respectively. 
The background in this channel 
originates from $\EEWW$, $\EEZZ$ and $\EQQG$ final states, 
and from triple gauge boson production 
processes. Events of the process $\mathrm{ZH\to q\bar{q} ZZ}$ 
constitute 9\% and 5\% of the signal
in the selected sample for $\MH$ = 150 and 180$\GeV$, respectively.

\subsection{Combined Results}
Table~\ref{table:higgs_mass} summarizes the 
statistical accuracy on the determination of $\MH$
for the different final states and their combination.
It should be noted that considerable overlap exists in the
selected samples of the $\zhllqq$ and $\zhllww$ channels and
of the $\zhqqqq$ and  $\zhqqww$ channels.  
Hence, the combination is performed only 
for the non-overlapping topologies which gives 
a minimal combined error on the Higgs boson mass. This is done
using the formula:
\begin{displaymath}
\frac{1}{\Delta^2(\MH)} = \sum_i{\frac{1}{\Delta_i^2(\MH)}},
\end{displaymath} 
where $\Delta$ is the combined error, whereas $\Delta_i$ is the 
error obtained in the $i^{th}$ channel. 

\begin{table}[ht]
\begin{center}
\begin{tabular}{|l|c|c|c|}
\hline 
                          & \multicolumn{3}{c|}{$\Delta(\MH)$ in$\MeV$}    \\ 
\cline{2-4} 
   Decay mode \hfill      &         120  &             150  &             180  \\ 
\hline
   $\zhllqq$ &          $\phantom{0}$85  &             100  & $\phantom{0}$--  \\
   $\zhqqqq$ &          $\phantom{0}$45  &             170  & $\phantom{0}$--  \\
   $\zhllww$ &          $\phantom{0}$--  & $\phantom{0}$90  & $\phantom{0}$80  \\
   $\zhqqww$ &          $\phantom{0}$--  &             100  &         150      \\
\hline
   Combined  &          $\phantom{0}$40  & $\phantom{0}$65  & $\phantom{0}$70  \\
\hline
\end{tabular}
\caption{Uncertainties on the  determination  of the Higgs boson
         mass for $\MH$ =
         120, 150 and 180$\GeV$. 
         The $\zhllww$ and $\zhqqww$ channels are used for the combination
         at $\MH$ = 150$\GeV$.
         } 
\label{table:higgs_mass}
\end{center}
\end{table}

\section{Beam Related Systematic Effects}

We have investigated the effect of a 
bias in the beam energy measurement, of the
beam energy spread and of an
uncertainty in the differential luminosity spectrum
on the measurement of the Higgs boson mass.

The impact of a bias
in the beam 
energy measurement is estimated
by generating signal samples with
both positron and electron beam
energies shifted with respect to the nominal value
of $\sqrt s$/2. These shifts are varied from -100$\MeV$ to 100$\MeV$ in 
25$\MeV$ steps. Since in the kinematical fit the energy is constrained
to the nominal value,
$\sqrt s$= 350$\GeV$, the shift in the beam 
energy is expected to result in a shift in 
the measured Higgs boson mass. 

\begin{figure}[h]
\begin{center}
\includegraphics*[width=0.6\textwidth]{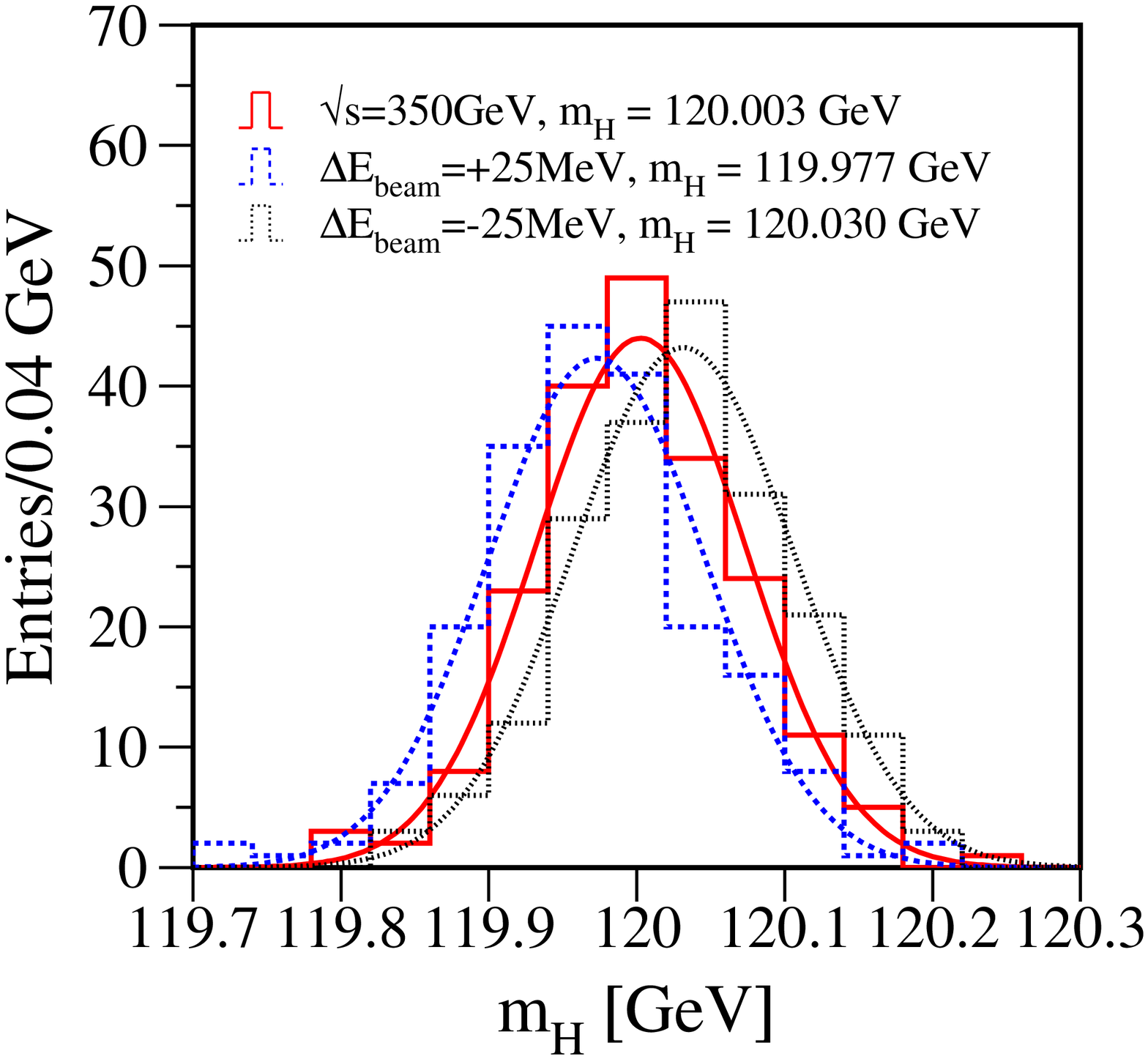}
\caption{The spectrum of the fitted values of the Higgs 
boson mass as obtained from 200 independent signal 
samples for the case when 
both electron and positron beam energies are overestimated 
by 25$\MeV$ (dotted histogram), 
when they are underestimated by 25$\MeV$ (dashed histogram) 
and when no shifts are introduced to the beam
energies (solid histogram).
\label{fig:beam_error}}
\vspace{-2mm}
\includegraphics*[width=0.6\textwidth]{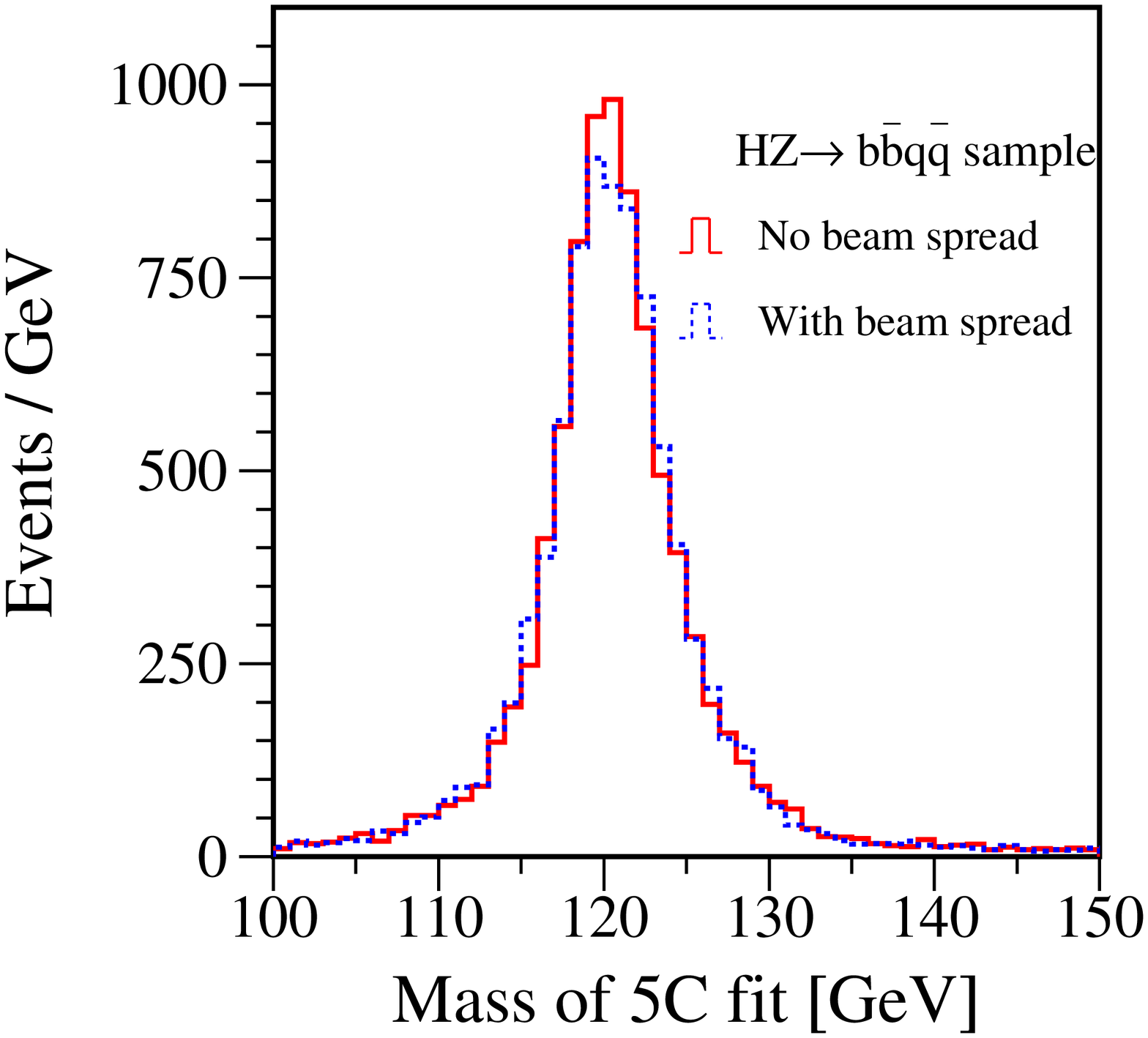}
\caption{Reconstructed Higgs boson mass spectrum in the 
sample of the $\zhqqqq$ events for the case of monochromatic
beams (solid histogram) and for the case of 1\% Gaussian 
energy spread for both electron and positron beams (dashed histogram). 
\label{fig:beam_spread}}
\end{center}
\end{figure}

As an example Figure~\ref{fig:beam_error} 
shows the distributions of fitted values of $\MH$
in the $\zhllqq$ channel for shifts in the 
beam energies of $+$25$\MeV$, zero$\MeV$ and $-$25$\MeV$. 
In each of the three considered cases the distribution
of  $\MH$
is obtained from 200 statistically independent signal samples.
The shift obtained in the fit of $\MH$ corresponds roughly to the
shift of the beam energy with opposite sign.
In the range of beam energy shifts from  -100 to 100$\MeV$
the shift in the Higgs boson mass is found to 
depend linearly on the shift in the beam energy:
\begin{equation}
\delta \MH = -\alpha \cdot \delta \rm{E_b},  
\end {equation}
with $\alpha =$ 0.85 for the  $\zhqqqq$ channel,
0.80 for the $\zhqqww$ channel,
and 
1.04 for the $\zhllqq$ and $\zhllww$ channels.
Hence, in order to keep the systematic bias in $\MH$ well
below its statistical error, the 
beam energy measurement
must be controlled with a precision better than 10$^{-4}$.

To estimate the impact of a beam energy spread, a Gaussian distribution 
of the beam energy has been used for the generation of signal
events. As an example, Figure~\ref{fig:beam_spread} 
shows the reconstructed Higgs boson mass spectrum for a sample of 
$\zhqqqq$ events for a 
1\% energy spread 
for both electron and positron beams
and the same distribution
for a fix  beam energy
of $\sqrt s$/2.
The inclusion of this beam energy spread slightly broadens
the mass distribution and degrades the
precision obtained  for $\MH$ from 45$\MeV$ to 50$\MeV$
in the $\zhqqqq$ channel and from 85$\MeV$ to 90$\MeV$ in the 
$\zhllqq$ channel.
For the TESLA machine the expected energy spread amounts to 
0.15\% for the electron beam and 0.03\% for the positron beam~\cite{tdr}. 
For these values of the beam energy spread, no significant 
degradation of the precision in 
the Higgs boson mass measurement is observed.

\begin{figure}[h]
\begin{center}
\includegraphics*[width=0.6\textwidth]{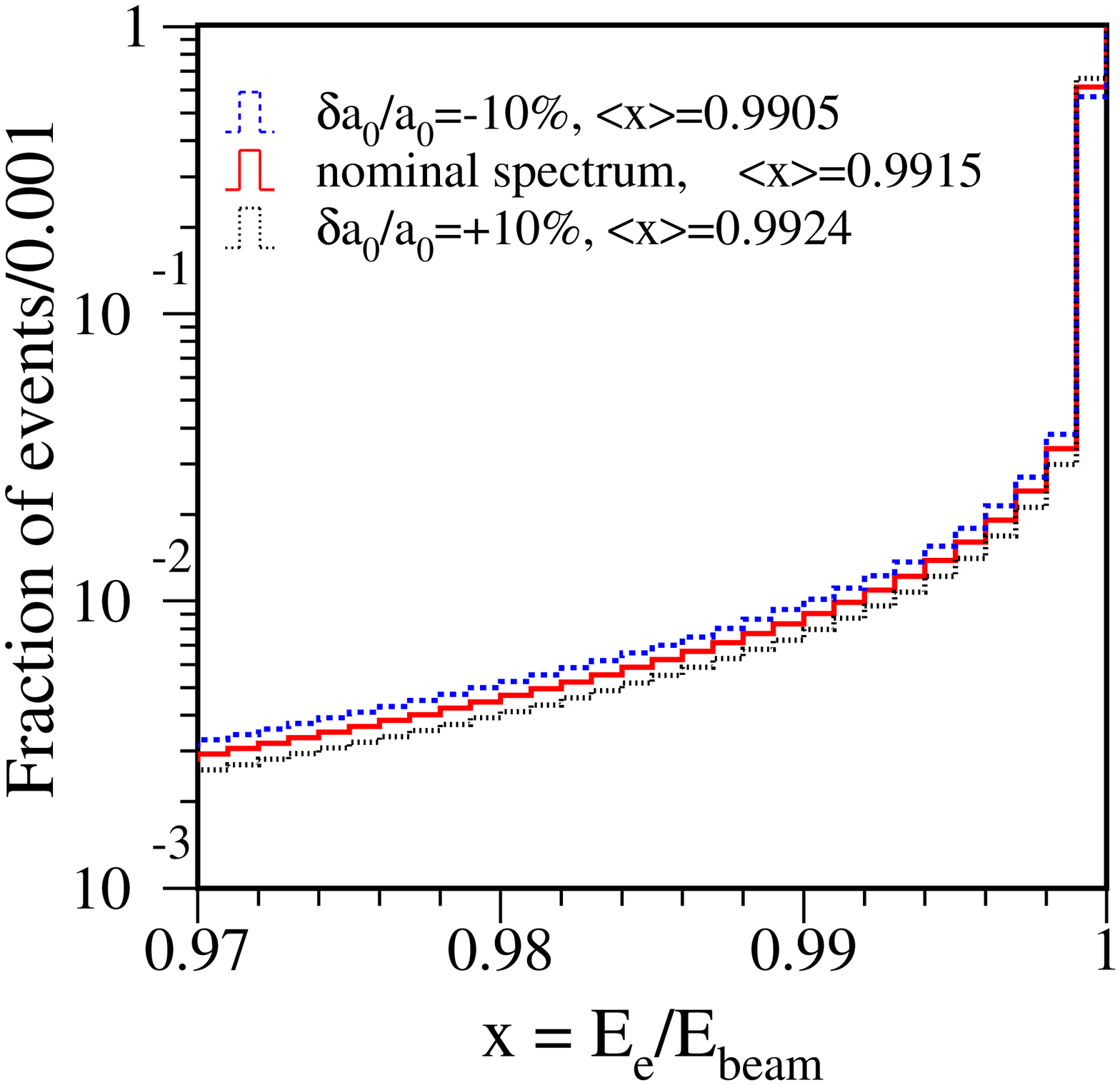}
\caption{The beam energy spectrum after beamstrahlung 
for nominal parameters $a_i$  at $\sqrt s =$ 350$\GeV$
(solid histogram) and for the cases when the parameter $a_0$ 
is shifted from its nominal value by -10\% (dashed histogram) and +10\% 
(dotted histogram). 
E$_{\rm{e}}$ is the energy of the colliding particle including beamstrahlung 
and E$_{\rm{beam}} $ is the nominal beam energy.
\label{fig:circe}}
\includegraphics*[width=0.6\textwidth]{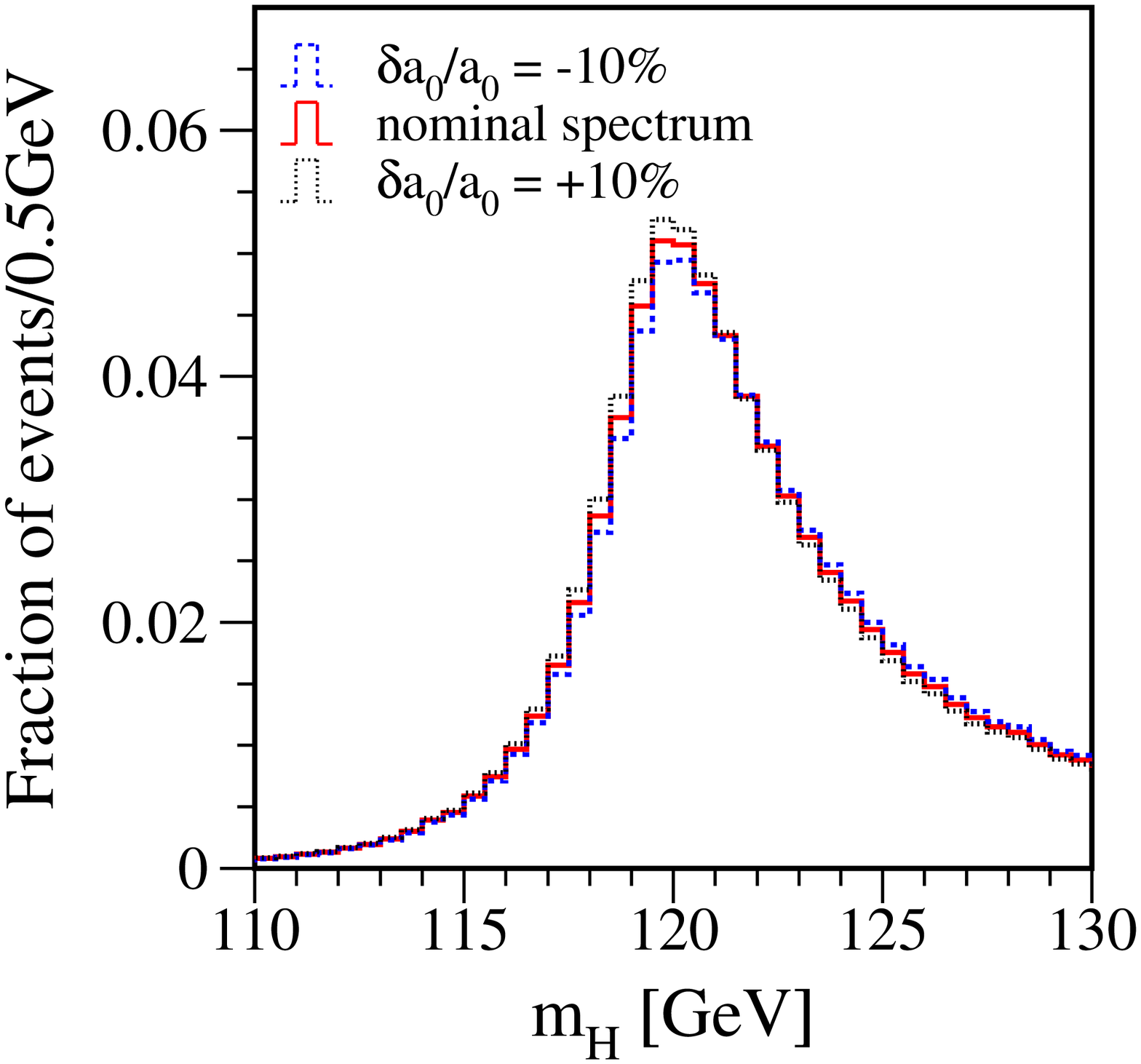}
\caption{The reconstructed Higgs boson mass spectrum 
in the sample of $\zhllqq$ events
for nominal parameters $a_i$ (see text) at 350$\GeV$ centre-of-mass energy 
(solid histogram) and for the cases when parameter $a_0$ 
is shifted from its nominal value 
by -10\% (dashed histogram) and +10\% (dotted histogram).
\label{fig:bstrahl}}
\end{center}
\end{figure}

The energy spectra of the colliding electrons and positrons at a high energy linear 
collider will be significantly affected
by photon radiation of the particles in one bunch in the coherent 
field of the opposite bunch. This effect is referred to as beamstrahlung.
The program CIRCE provides a fast simulation of the  beamstrahlung
under the assumptions that
the beamstrahlung in 
the two beams is equal and uncorrelated between the beams. The spectrum 
is parameterized according to
\begin{displaymath}
f(x) = a_0\delta(1-x)+a_1x^{a_2}(1-x)^{a_3},
\end{displaymath}
where $x$ is the ratio between the energies of the colliding electron and positron 
and
the initial energy of the undisrupted beam. The parameters $a_i$ 
depend on the operational  
conditions of the linear collider. The normalization 
condition, $\int{f(x)dx=1}$,
fixes one of these parameters, leaving only three of them independent. 
The default parameters for the TESLA machine 
operated at the centre-of-mass energy of 350$\GeV$ are:
\begin{displaymath}
a_0=0.55,\hspace{2mm}a_1=0.59,\hspace{2mm}a_2=20.3,\hspace{2mm}a_3=-0.63.
\end{displaymath}
It has been shown that from the analysis of 
the acollinearity spectrum of Bhabha
scattering
events, the parameters $a_i$ can be determined with a precision of about 
1\%~\cite{lumi}.  
To visualize the effect of the uncertainty in the determination of the parameters 
$a_i$,  
the beam energy spectra
are shown
 in Figure~\ref{fig:circe} 
for nominal values of the parameters $a_i$ and for  
the parameter $a_0$ shifted by $\pm$10\% from its nominal value.  
Figure~\ref{fig:bstrahl} presents the corresponding Higgs boson mass 
spectra a the sample of $\zhllqq$ events. 
An uncertainty of 10\% in the determination of the parameters $a_0$ 
results in a systematic uncertainty of about 10$\MeV$ 
on the Higgs boson mass
in the $\zhllqq$ and $\zhqqqq$ channels. The same result is obtained for
the other parameters.
The uncertainty is reduced to  about 1$\MeV$ if the parameters $a_i$ are measured
with an accuracy of 1\%. 
The same result is obtained for the
study of the $\zhllww$ and $\zhqqww$ channels.

\section{Conclusion}
The potential of the future linear $\EE$ collider for the 
measurement of the Higgs boson mass is evaluated.
Assuming an integrated luminosity of 500 fb$^{-1}$,
 the Higgs boson
mass can be measured with a statistical accuracy ranging from 
40$\MeV$ to 70$\MeV$ for 
$\MH$ between 120$\GeV$ and 180$\GeV$. In order to keep the systematic 
uncertainty due to a bias of the beam energy measurement well below 
the statistical uncertainty, the beam energy
measurement has to be controlled with a  
precision better than 10$^{-4}$. Under operational conditions envisaged
for the TESLA machine, the beam energy spread and uncertainty in the 
differential luminosity spectrum are found to have negligible effect on the 
Higgs boson mass measurement.

\section{Acknowledgments}
We would like to thank Prof. K. Desch for many helpful discussions
and his continuous interest and support. 

\newpage

\vspace{-1cm}

\end{document}